\documentclass[12pt]{iopart}
\usepackage{iopams}
\usepackage{graphicx}
\usepackage{color}

\begin{document}

\newcommand{\REV}[1]{\textbf{\color{red}#1}}
\newcommand{\BLUE}[1]{\textbf{\color{blue}#1}}
\newcommand{\GREEN}[1]{\textbf{\color{green}#1}}
\newcommand{\rev}[1]{{\color{red}#1}}

\title[Binary BEC in a confining potential]{Binary mixtures of condensates
in  generic confining potentials}

\author{P. Facchi $^{1,2}$, G. Florio$^{3,2}$, S. Pascazio$^{3,2}$, F. V. Pepe$^{3,2}$}

\address{$^1$Dipartimento di Matematica and MECENAS, Universit\`a di Bari, I-70125 Bari, Italy}
\address{$^2$INFN, Sezione di Bari, I-70126 Bari, Italy}
\address{$^3$Dipartimento di Fisica and MECENAS, Universit\`a di Bari, I-70126 Bari, Italy}

\begin{abstract}
We study a binary mixture of Bose-Einstein condensates, confined
in a generic potential, in the Thomas-Fermi approximation. We
search for the zero-temperature ground state of the system, both in
the case of fixed numbers of particles and fixed
chemical potentials.

\end{abstract}


\vspace{2pc}

\submitto{\JPA}

\section{Introduction}

Binary mixtures of Bose-Einstein condensates are of great interest
due to their complex dynamical features and their role in the
emergence of macroscopic quantum phenomena. Mixtures
are usually made up of two species, that can also be hyperfine states of
the same alkali atom \cite{alkali}. They generally display repulsive
self-interaction and are confined by different potentials.
Depending on the inter-species interaction, two classes of stable
configurations are possible: mixed and separated. The latter are
more interesting, since they allow the observation of phenomena
such as symmetry breaking, e.g.\ in harmonic potentials, and
macroscopic quantum tunnelling \cite{kasamatsu,trippenbach}.
Binary mixtures in harmonic traps have been investigated in a
number of interesting experiments \cite{myatt,hall,firenze}.

Different approaches are possible in order to study the ground
state of these systems. The binary mixture of two species of
bosons can be rigorously described in a second-quantization
formalism \cite{ohberg}. However, if the number of particles in
the condensate is very large compared to the number of particles
in the excited states, the fields associated to the two species
can be treated as classical wave functions. This approach leads to
the Gross-Pitaevskij equations \cite{book}, that are obtained by
minimizing the zero-temperature grand-canonical energy of the
system. The ground state of the system can be thus determined by
solving the Gross-Pitaevskij equation
\cite{lieb,timmermans,ao1,graham}, or equivalently by analytically
or numerically minimizing the grand-canonical energy functional
\cite{trippenbach,kasamatsu,pu}. Analytical results are obtained
only in particular cases, such as confinement by a hard wall trap \cite{wall}, harmonic or lattice potentials \cite{harmonic} and  axisymmetric traps \cite{axis}. A simplified approach is often used,
based on the Thomas-Fermi (TF) approximation, that consists in
neglecting the kinetic energy with respect to the self- and
inter-species interaction energies \cite{book}. This reduces the
problem of finding the ground state of the binary mixture of
condensates to a classical problem, related to the stability of a
system of two interacting fluids.

In this article we shall investigate this problem by adopting the
following approach: given a system of two interacting condensates,
confined in a generic external potential (that can be different
for the two species), we will find \textit{general} tools to
determine the ground state of such system in the TF approximation.  The article has the
following structure. In \Sref{gpe-tfa} we set up the problem and
introduce notation. In \Sref{mix-sep} we find a threshold value of
the inter-species interaction parameter, above which mixed
configurations cannot be the ground state of the system. In
\Sref{sep-opt} we establish those conditions that determine which
one of the possible separated configurations is the ground state.
We conclude with an outlook in \Sref{concl}. Throughout this
article, both cases of i) fixed numbers of particles and ii) fixed
chemical potentials will be considered.

\section{Gross-Pitaevskij equations and Thomas-Fermi
solutions}\label{gpe-tfa}

We consider a system made up of two species of indistinguishable
particles, labelled 1 and 2, confined by generally different
external potentials $V_1(x)$ and $V_2(x)$. Self interaction and
inter-species interaction are assumed to be repulsive. An example
of such a system is a mixture of alkali atoms in two different
hyperfine states \cite{myatt,hall}. The two subsystems are
described in a quantum field theoretical framework, by associating
to each species the field operators $\hat{\psi}_1(x)$ and
$\hat{\psi}_2(x)$. However, since we are searching for the
zero-temperature ground state, we assume that all particles
condense in the same wave function, and thus apply a Bogolubov
shift \cite{yukalov} and treat $\psi_1(x)$ and $\psi_2(x)$ as
classical fields, normalized to the average numbers of particles $N_1$ and $N_2$.
The grand-canonical energy
functional is
\begin{equation}
\mathcal{E} = \mathcal{T} + \mathcal{U} -\mu_1 (\mathcal{N}_1 - N_1) - \mu_2 (\mathcal{N}_2-N_2),
\label{engcpsi}
\end{equation}
where
\begin{eqnarray}
\fl\qquad\quad \mathcal{T} &=& \int
\left(\frac{\hbar^2}{2m_1}|\nabla\psi_1|^2
+\frac{\hbar^2}{2m_2}|\nabla\psi_2|^2
\right) d^m x, \nonumber\\
\fl \qquad\quad \mathcal{U} &=& \int \left(V_1|\psi_1|^2+V_2|\psi_2|^2+
\frac{U_{11}}{2}|\psi_1|^4+\frac{U_{22}}{2}|\psi_2|^4
+U_{12}|\psi_1|^2|\psi_2|^2 \right) d^{m}x,
\nonumber\\
\fl \qquad \mathcal{N}_k&=& \mathcal{N}(|\psi_k|^2)=\int |\psi_k(x)|^2 \, d^{m}x ,\qquad
k=1,2,
\label{engcpsi1}
\end{eqnarray}
and $m$ is the dimension of the system. By requiring that the
energy is stationary, one obtains the coupled Gross-Pitaevskij (GP)
equations
\begin{eqnarray}\label{gpe}
\left(-\frac{\hbar^2}{2m_1}\Delta +V_1(x)+U_{11}|\psi_1(x)|^2+U_{12}|\psi_2(x)|^2\right)\psi_1(x)=\mu_1\psi_1(x) \, ,
\\
\label{gpe2}
\left(-\frac{\hbar^2}{2m_2}\Delta+V_2(x)+U_{22}|\psi_2(x)|^2+U_{12}|\psi_1(x)|^2\right)\psi_2(x)=\mu_2\psi_2(x)\,.
\end{eqnarray}
In the above equations $U_{11}$ and $U_{22}$ are the
self-interaction parameters between atoms of the same species,
while $U_{12}$ is associated to inter-species interaction. Each of
these parameters is assumed to be positive, since we are considering
repulsive interactions. The solutions of \eref{gpe}-\eref{gpe2}
depend on the value of the chemical potentials $\mu_1$ and
$\mu_2$, which are Lagrange multipliers. If $\mu_1$ and $\mu_2$
are fixed, the average particle numbers are free to vary. If, on
the other hand, the particle numbers $N_1$ and $N_2$ are fixed,
the chemical potentials are chosen in such a way that the wave
functions satisfy the normalization constraints
$\mathcal{N}_k=N_k$, for $k=1,2$.

For the sake of simplicity, our analysis will be focused on
one-dimensional systems, with the main results generalizable to
higher dimensions. Moreover, it will  be assumed that the
potentials be  continuously differentiable, $V_k\in C^1
(\mathbb{R})$. This class of potentials schematizes very well those used in trapping cold atoms.

The Thomas-Fermi (TF) approximation, which will be applied in the
following, consists in neglecting the kinetic energy contribution
$\mathcal{T}$ to the energy functional \eref{engcpsi}. This approximation is justified
if the number of particles is sufficiently high, since the
self-energetic parts in Eq.\ \eref{engcpsi} are respectively
$\Or(N_1^2)$ and $\Or(N_2^2)$, while the kinetic energy is
$\Or(N_1^{3/2})+\Or(N_2^{3/2})$ \cite{book}. As a consequence of
the TF approximation, the grand-canonical energy becomes dependent
only on the densities $\rho_1(x)=|\psi_1(x)|^2$ and
$\rho_2(x)=|\psi_2(x)|^2$, and will be indicated in the following
as $\mathcal{E}_{\mathrm{TF}}(\rho_1,\rho_2)$.

Notice that, without loss of generality,  one can reduce the
analysis to the particular case $U_{11}=U_{22}=1$. Indeed, by
the scaling
\begin{equation}
\label{eq:scaling}
\fl\qquad \rho_k \to \rho_k/\sqrt{U_{kk}}, \quad N_k \to N_k/\sqrt{U_{kk}}, \quad V_k \to V_k \sqrt{U_{kk}},
\quad \mu_k \to \mu_k \sqrt{U_{kk}}
\end{equation}
one gets
\begin{equation}
\mathcal{E}_{\mathrm{TF}}(\rho_1,\rho_2) =
\mathcal{U}(\rho_1,\rho_2) -
\mu_1\left(\mathcal{N}(\rho_1)-N_1\right)
-\mu_2\left(\mathcal{N}(\rho_2)-N_2\right), \label{eq:TFenergy}
\end{equation}
with
\begin{equation}
\mathcal{U}(\rho_1,\rho_2)=\frac{1}{2}\int \left( \rho_1^2 +
\rho_2^2 + 2 \alpha \rho_1 \rho_2 \right) dx + \int\left( V_1
\rho_1 + V_2 \rho_2 \right) dx, \label{eq:intenergy}
\end{equation}
and
\begin{equation}
\label{eq:alphadef}
\alpha=\frac{U_{12}}{\sqrt{U_{11}U_{22}}}.
\end{equation}
Incidentally, notice that the above reduction to a single
parameter $\alpha$ applies also to the full energy functional
(\ref{engcpsi}), by scaling also the masses $m_k \to
m_k/\sqrt{U_{kk}}$.

The critical points
of the Thomas-Fermi grand-canonical energy functional are
the solutions to the algebraic equations,
\begin{equation}\label{gpeTFA1}
\rho_1(x) + \alpha \rho_2(x) + V_1(x)=\mu_1, \qquad
\rho_2(x) + \alpha \rho_1(x) + V_2(x)=\mu_2
\end{equation}
and will be called the TF density profiles. Moreover, for fixed
particle numbers $N_k$, they are supplemented by the normalization
conditions
\begin{equation}
\int \rho_k \, dx = N_k, \qquad (k=1,2),
\label{eq:normcond}
\end{equation}
which fix the values of the chemical potentials $\mu_k$.

In the following the supports of the TF
densities $\rho_k$ will be denoted  by
$S_k$. By assuming that $\alpha\neq 1$, in $S_{12}=S_1\cap S_2$,
where both condensates are present, the TF density profiles are
\begin{equation}\label{TFAmix1}
\fl \quad \rho_1(x)=\frac{\mu_1 - V_1(x) -\alpha \left(\mu_2- V_2(x)\right)}{1-\alpha^2} \, ,
\quad \rho_2(x)=\frac{\mu_2 - V_2(x) -\alpha \left(\mu_1- V_1(x)\right)}{1-\alpha^2} \, .
\end{equation}

In the regions $S_{11}=S_1-S_2$ and $S_{22}=S_2-S_1$, occupied by only one of
the two species, the solutions are respectively
\begin{equation}\label{TFA1}
\rho_1(x)=\mu_1-V_1(x), \qquad \rho_2(x)\equiv 0,
\end{equation}
and
\begin{equation}\label{TFA2}
\rho_2(x)=\mu_2-V_2(x) , \qquad \rho_1(x)\equiv 0 .
\end{equation}
The TF density profiles \eref{TFAmix1}--\eref{TFA2} are defined
independently of the dimensionality of the system.

Notice that the TF equations \eref{gpeTFA1} uniquely determine the
functional dependence of the densities at a point $x$ on the
external potentials at the same point, the chemical potentials and
the interaction parameters, \emph{once} the supports $S_1$ and
$S_2$ are given. On the other hand large freedom is left in the
choice of the supports of the density profiles, for which
uniqueness fails. Thus, extremely irregular configurations can be
solutions of the TF equations. Among all possible solutions, one
should pick up the minimizers.

The rest of this paper will be devoted to deriving general
rules for finding the minimizing configuration of the supports, in
order to determine the ground state of the system, both if the
numbers of particles or the chemical potentials are fixed.

\section{Mixed vs separated configurations}\label{mix-sep}

The configurations of the binary mixture can be divided in two
fundamental parts: separated and mixed. The TF densities are
\textit{mixed}  in $S_{12}=S_1\cap S_2$, where both species are
present, and are  \textit{separated} in $S_{11} \cup S_{22}= S_1
\cup S_2 - S_1\cap S_2$, where only one species is present at one
time. A configuration is said to be separated if it does not
contain mixed parts, and mixed otherwise.
In this Section we will show that
\begin{equation}\label{u12lim}
\alpha^{\mathrm{th}}=1 \qquad (U_{12}^{\mathrm{th}}=\sqrt{U_{11}U_{22}})
\end{equation}
plays the role of a threshold value, above which separated
configurations become energetically favored, both in the case of
 i) fixed numbers of particles and ii) fixed chemical potentials.
This threshold holds independently of the particular external
potentials $V_{k}(x)$. We proceed by treating separately cases i)
and ii).

\subsection{Solutions are confined}

We will first prove that under the assumption that the $C^{1}$
potentials are confining, that is
\begin{equation}
V_k(x) \to +\infty, \qquad \mathrm{for} \quad |x|\to\infty,
\end{equation}
with $k=1,2$, all TF density profiles are compactly supported. We
will see that this is a straight consequence of the positivity of
the densities
\begin{equation}
\rho_k(x) \geq 0.
\end{equation}
We will prove that the supports $S_k$ are bounded, by separately
considering the sets with separated phases, $S_{11}=S_1-S_2$ and
$S_{22}=S_2-S_1$, and that with mixed phases, $S_{12}=S_1\cap
S_2$. By requiring that the solutions (\ref{TFA1}) and
(\ref{TFA2}) be nonnegative we get
\begin{eqnarray}
S_{11}\subset\{x\in\mathbb{R} \,|\, V_1(x)\leq \mu_1 \} = V_1^{-1}(-\infty,\mu_1],
\nonumber\\
S_{22}\subset\{x\in\mathbb{R} \,|\, V_2(x)\leq \mu_2 \} = V_2^{-1}(-\infty,\mu_2],
\label{eq:musep}
\end{eqnarray}
which are bounded by hypothesis. On the other hand, from
(\ref{TFAmix1}) we get that every point $x\in S_{12}$ satisfies
the conditions
\begin{equation}
\fl \qquad \frac{\mu_1 - V_1(x) -\alpha \left(\mu_2- V_2(x)\right)}{1-\alpha^2} \geq 0 \, ,
\qquad \frac{\mu_2 - V_2(x) -\alpha \left(\mu_1- V_1(x)\right)}{1-\alpha^2}\geq 0.
\end{equation}
They are easily proved to be equivalent to
\begin{equation}
\fl\qquad \min\{\alpha,\alpha^{-1}\} \left(\mu_2 - V_2(x)\right)
\leq \left(\mu_1-V_1(x)\right)\leq \max\{\alpha,\alpha^{-1}\}
\left(\mu_2 - V_2(x)\right),
\end{equation}
which in turn imply that
\begin{equation}
S_{12}\subset V_1^{-1}(-\infty,\mu_1] \cap V_2^{-1}(-\infty,\mu_2],
\label{eq:mumix}
\end{equation}
so that $S_{12}$ is compact. As a consequence $S_1 = S_{11}\cup
S_{12}$ and $S_2=S_{22}\cup S_{12}$ are compact.

\subsection{Fixed numbers of particles}

If the numbers of particles $N_1$ and $N_2$ are kept fixed, the
chemical potentials are functionally dependent on the density
profiles, since they have to be tuned in order to preserve the
normalization conditions~(\ref{eq:normcond}).
The search for the zero-temperature ground state of the system
reduces to the minimization of the TF grand-canonical energy
functional~(\ref{eq:TFenergy}), that evaluated at the TF solutions reduces
to the internal energy functional~(\ref{eq:intenergy}).

\subsubsection{Square well.}

A simple lemma will be now introduced (see e.g. \cite{ao2}).
Consider a binary mixture confined in an infinite square well,
corresponding to a bounded interval $S=[a,b]$ (with $b>a$) of the
real axis. Let $|S|$ be the (finite) length of the well. Since in
this case $V_k(x)\equiv 0$ and the TF density profiles are flat,
$\rho_k(x) = N_k/|S|$, the internal energy of the completely mixed
configuration in $S$ is
\begin{equation}
\mathcal{U}_{\mathrm{m}}
=\frac{1}{2 |S|}\left(N_1^2+
N_2^2+2 \alpha N_1N_2\right).
\end{equation}
On the other hand, a separated configuration with $N_1$ particles
of the first species in a subset $S_1\subset S$ and $N_2$
particles of the second in $S_2=S-S_1$ has densities
$\rho_k(x)=N_k/|S_k|$ and internal energy
\begin{equation}
\mathcal{U}_{\mathrm{s}}(|S_1|)
=\frac{1}{2}\frac{N_1^2}{|S_1|}+\frac{1}{2}\frac{N_2^2}{|S-S_1|},
\end{equation}
which is in fact a function of the length $|S_1|$. The minimum of
$\mathcal{U}_{\mathrm{s}}$ is attained for supports $\bar{S}_1$
and $\bar{S}_2=S-\bar{S}_1$ such that
\begin{equation}\label{partvol}
\frac{N_1}{|\bar{S}_1|}=
\frac{N_2}{|\bar{S}_2|} .
\end{equation}
Condition \eref{partvol}  can be also expressed in terms of
the densities in the separated configuration:
\begin{equation}\label{densratio}
\bar{\rho}_1=\bar{\rho}_2.
\end{equation}
Since the value of the internal energy at
\eref{partvol} is
\begin{equation}\label{enintsqwS}
\bar{\mathcal{U}}_{\mathrm{s}}=
\frac{1}{2|S|}\left(N_1^2+ N_2^2+ 2 N_1N_2\right),
\end{equation}
the minimizing separated configurations are energetically favorite if
$\alpha\geq1$, while the mixed configurations are
less energetic than all separated configurations if
$\alpha<1$. For a binary mixture in a square
well, this proves the role of Eq.\ \eref{u12lim} as a
threshold value.

\subsubsection{Selection principle and regularity of  solutions.}
\label{sec:selprinc} Notice that the minimizers (\ref{densratio})
have a very high degeneracy, since every sets $\bar{S}_1$ whose
measures satisfy (\ref{partvol}) correspond to  possible TF
configurations. Among them, despite the regularity of the
potentials, there are extremely irregular configurations with
highly entangled supports and infinitely many points of discontinuity
(\emph{domain walls}). However, such a phenomenon is a consequence
of the TF approximation that, by neglecting the kinetic energy
part $\mathcal{T}$ in (\ref{engcpsi}), is also deprived of its
regularizing effect on the densities. The kinetic term favors
smooth density profiles. Indeed, due to the presence of the
kinetic energy, the grand-canonical energy functional
(\ref{engcpsi}) is defined on functions with square integrable
(distribution) derivatives, and TF solutions are approximations thereof. In particular, in the 1-dimensional situation, a domain wall
is a discontinuous approximation of a (absolutely) continuous
function that changes between two values in a very short transition
region with a large derivative.

Thus, each domain wall of the TF solution would correspond to an additional
cost, in terms of kinetic energy of the true solution, and the above-mentioned degeneracy would be lifted: among the TF degenerate
minimizers, $\mathcal{T}$ would select those one(s) with the minimum
number of domain walls.
 In the following we will make use of this \emph{selection principle} and, in particular, we will only consider densities  in the class of piecewise  differentiable functions, $\rho_k \in \tilde{C}^1$. That means that there is a finite subdivision of $S_k$ such that the restriction of $\rho_k$ to each subinterval $[t_j,t_{j+1}]$ is continuously differentiable.
 Incidentally, notice that  the potentials themselves  can be assumed to be piecewise differentiable, without modifying our results.

Going back to the square-well case, by the selection principle, for $\alpha\geq 1$, we end up with only two degenerate minimizers: one with $S_1= [a,c]$,  where $c=(a N_2 + b N_1)/(N_1+N_2)$,
 and the other with $S_1=[d, b]$, where $d=(a N_1+b N_2)/(N_1+N_2)$. Both configurations are separated and have a single domain wall.

\subsubsection{Generic potential.}
We now extend the above result to the case in which the mixture is
not confined in a square well, but rather by generic continuously differentiable
confining potentials $V_{k}(x)$ with $k=1,2$.
We will prove that, if
$\alpha\geq1$, the ground state of the system
cannot be a mixed configuration, and thus we can restrict our
attention to the separated ones.

Let the TF densities have a  mixed configuration in $S_{12}=S_{1}\cap S_2$, given by Eq.~(\ref{TFAmix1}).
Since the densities  $\rho_k$ are assumed to be piecewise
continuously differentiable, and their supports are compact, $S_{12}$ is the union of a finite number of compact intervals in which the $\rho_k$ are $C^1$.
Choose a segment
$\omega=[x_0,x_1]$ of length $|\omega|=\varepsilon>0$ included in some of those intervals. The interval $\omega$
contains
\begin{equation}
n_j=\int_{\omega} \rho_k(x)\, dx = \varepsilon \langle\rho_k\rangle ,
\qquad (k=1,2)
\end{equation}
particles, where $\langle\cdot\rangle$ denotes the average on $\omega$. Since the potentials are $C^1$, we can express them in $\omega$ as
\begin{equation}\label{Vder}
V_{k}(x) = V_{k}(x_0)+V'_k (\xi_k(x))(x-x_0) , \qquad (k=1,2)
\end{equation}
for some $\xi_{k}(x)\in \omega$.

Since the  $\rho_k$ are continuous in $\omega$, two points
$\bar{x}_k$ exist in this segment, in which the functions equal
their averages:
\begin{equation}\label{aver}
\rho_k(\bar{x}_k)=\langle\rho_k\rangle .
\end{equation}
Moreover, since the density functions are continuously
differentiable in $[x_0,x_1]$, taking into account \eref{aver} we
can express them in each point of the segment as
\begin{equation}\label{rhoder}
\rho_{k}(x) =\langle\rho_{k}\rangle+ \rho'_{k}(\eta_{k}(x))
(x-\bar{x}_{k}),
\end{equation}
for some $\eta_{k}(x)\in \omega$. The first derivatives
appearing in \eref{Vder}-\eref{rhoder} are all bounded in $\omega$.
Thus, the internal energy of particles in the set
$\omega$ reads
\begin{equation}\label{enintsegM}
\fl \quad u_{\mathrm{m}} =
\varepsilon\left[
\frac{1}{2}\langle\rho_1\rangle^2
+\frac{1}{2}\langle\rho_2\rangle^2
+\alpha\langle\rho_1\rangle\langle\rho_2\rangle
+V_1(x_0)\langle\rho_1\rangle+V_2(x_0)\langle\rho_2\rangle
\right]+\Or(\varepsilon^2).
\end{equation}
We now divide  $\omega$ in two
subintervals $\omega_1=[x_0,y]$ and $\omega_2=[y,x_1]$:
\begin{equation}
\omega=\omega_1\cup\omega_2.
\end{equation}
and replace the TF mixed densities in $\omega$ with flat and
separated density profiles, preserving the particle numbers
\begin{eqnarray}\label{dsepseg1}
\bar{\rho}_1=\frac{n_1}{|\omega_1|} \qquad \mathrm{for } \quad
x\in\omega_1,
\\ \label{dsepseg2} \bar{\rho}_2=\frac{n_2}{|\omega_2|} \qquad
\mathrm{for } \quad x\in\omega_2.
\end{eqnarray}
As a rule in choosing the bipartition of $\omega$, we assume that
the stationarity condition \eref{partvol} for the internal energy
in an infinite potential well is satisfied
\begin{equation}
\frac{|\omega_2|}{|\omega_1|}=\frac{n_2}{n_1}.
\end{equation}
Taking into account the result \eref{enintsqwS}, concerning the
self-interaction and inter-species interaction parts, the
potential energy of the set $\omega$ with separated densities
\eref{dsepseg1}-\eref{dsepseg2} can be expressed, after a
straightforward manipulation, as
\begin{equation}
\fl \quad u_{\mathrm{s}}  =  \varepsilon
\left[\frac{1}{2}\langle\rho_1\rangle^2
+\frac{1}{2}\langle\rho_2\rangle^2 +
\langle\rho_1\rangle\langle\rho_2\rangle
+V_1(x_0)\langle\rho_1\rangle+V_2(x_0)\langle\rho_2\rangle
\right]
+\Or(\varepsilon^2) .
\end{equation}
The net change in the total potential energy, due to the
replacement of the mixed densities in $[x_0,x_1]$ with the
separated ones, is
\begin{equation}\label{diffpot}
\delta\mathcal{U}= u_{\mathrm{s}}-u_{\mathrm{m}}=\varepsilon
\left(1-\alpha\right) \langle\rho_1\rangle\langle\rho_2\rangle
+\Or(\varepsilon^2) \,.
\end{equation}
 For sufficiently small $\varepsilon$, the sign of
$\delta\mathcal{U}$ is determined by the first term in
\eref{diffpot}, unless $\alpha=1$.

If $\alpha>1$, the result $\delta\mathcal{U}<0$
implies that, given a point of a mixed configuration, there always exists a neighborhood in which one can construct a
 separated configuration with lower energy. Since $S_{12}$ is compact, we can find a finite subdivision $S_{12}=[x_0,x_1]\cup \dots \cup [x_{n-1},x_{n}]$, such that the above construction can be performed in each segment $[x_{j-1},x_j]$.
Thus, the minimizers are separated configurations, if the particle
numbers are fixed. Analogously, one can show
that if $\alpha<1$ the internal energy of a
separated configuration is always larger than the energy of a
mixed configuration with the same particle numbers. Thus, even in
the case of varying external potentials the value \eref{u12lim}
acts as a threshold between mixed and separated ground states.

\subsection{Fixed chemical potentials}

If the chemical potentials $\mu_1$ and $\mu_2$ are fixed, the
average particle numbers are free to vary. In order to find the
ground state of the system, one has to find the minimizers of the grand-canonical
energy $\mathcal{E}_{\mathrm{TF}}(\rho_1,\rho_2)$. As in
the case of fixed particle numbers, the elementary case of a
binary mixture in an infinite square well will be analyzed first.
Then, we will try and find general results in the case of
different piecewise continuously differentiable confining
potentials.

\subsubsection{Square well.}
It is clear from \eref{TFAmix1}--\eref{TFA2} that
fixing the chemical potentials corresponds to fixing the density
functions, with the choice of the supports leading to different
numbers of particles. We will neglect an inessential annoying constant in~(\ref{eq:TFenergy}) by setting $N_1=N_2=0$.

In the simple case of an infinite
square well $S$ with $V_1=V_2=0$ inside the well, the only
values of the chemical potentials that have physical meaning are
the positive ones, as it emerges from (\ref{eq:musep}) and (\ref{eq:mumix}). Let us first consider a completely mixed
configuration in the well with density profiles
\begin{equation}\label{tfam1}
\rho_1^{\mathrm{m}}=\frac{\mu_1-\alpha \mu_2}{1-\alpha^2} \,, \qquad \rho_2^{\mathrm{m}}=\frac{\mu_2 -\alpha \mu_1}{1-\alpha^2}.
\end{equation}
Such a configuration has a physical meaning if both densities are
non negative. Thus, if $\alpha<1$ the
numerators in \eref{tfam1} must be nonnegative,
while if $\alpha>1$ the numerators must be nonpositive. Densities in a mixed configuration are not defined at the threshold value $\alpha=1$.
 Conditions on the positivity of the densities set a
bound on the values of $\alpha$ which are compatible with the
chosen chemical potentials:
\begin{equation}
\alpha \notin (\alpha^{\mathrm{l}},\alpha^{\mathrm{u}}),
\end{equation}
where
\begin{equation}\label{lboundu12}
\alpha^{\mathrm{l}}:= \min\left\{\frac{\mu_1}{\mu_2},\frac{\mu_2}{\mu_1}\right\}\leq 1 \leq \alpha^{\mathrm{u}} :=
\max\left\{\frac{\mu_1}{\mu_2},\frac{\mu_2}{\mu_1}\right\}.
\end{equation}
The problem of nonphysical values of $\alpha$  does not arise in
the case
\begin{equation}\label{stat}
\mu_2=\mu_1,
\end{equation}
which will prove to be a very relevant physical situation. If
\eref{stat} holds, the boundaries \eref{lboundu12} coincide, and the densities are well defined for
all $\alpha$. Note that condition
\eref{stat} exactly corresponds to the minimum condition of the
internal energy in the separated phase, since, by taking into account
\eref{TFA1}-\eref{TFA2}, it implies
\begin{equation}
\rho_1^{\mathrm{s}}=\rho_2^{\mathrm{s}}.
\end{equation}
If the solutions \eref{tfam1} are plugged in the
definition of the grand canonical energy, it is possible to
express it in terms of interaction parameters and chemical
potentials
\begin{equation}\label{engcM}
\mathcal{E}_{\mathrm{m}} =
|S|\left[\frac{\mu_1^2+\mu_2^2-2\mu_1\mu_2\alpha}{2\left(
\alpha^2-1 \right)}\right] .
\end{equation}
If instead separated solutions are considered, with the first
condensate confined in a region $S_1$ and the second in
$S_2=S-S_1$, the grand-canonical energy is a function of
the length $|S_1|$ and reads
\begin{equation}\label{engcS}
\mathcal{E}_{\mathrm{s}}(|S_1|)=-|S_1|\frac{\mu_1^2}{2}
-|S-S_1|\frac{\mu_2^2}{2} \,.
\end{equation}
Since \eref{engcS} is linear in the length $|S_1|$, it is
clear that its minimum value is
\begin{equation}\label{engcSo}
\bar{\mathcal{E}}_{\mathrm{s}}=-|S|\max\left\{\frac{\mu_1^2}{2},
\frac{\mu_2^2}{2}\right\} \,.
\end{equation}
Hence, if $\mu_2<\mu_1$, the minimum of
the grand-canonical energy for separated configurations
corresponds to $S_1=S$, while if
$\mu_2>\mu_1$, it corresponds to
$S_2=S$. In both cases, the minimizer is in fact
a \textit{single condensate} configuration. Only if \eref{stat}
holds, separated configurations are allowed. Moreover, their
energy is stationary with respect to changes in the partition of
$S$.

The energy in the mixed configuration and that in the minimizing
separated configuration will be now compared in detail, as
$\alpha$ varies from zero to infinity, in the cases
$\mu_2 <\mu_1$ and
$\mu_2=\mu_1$, since if
$\mu_2>\mu_1$ the physical situation is
specular with respect to the first case. We remark that the energy
of each separated configuration is always independent of $\alpha$.

In the case $\mu_2<\mu_1$, if the
inter-species interaction is absent, $\alpha=0$, the mixed configuration is
favorite, since its energy $\mathcal{E}_\mathrm{m}^0$ reads
\begin{equation}
\mathcal{E}_{\mathrm{m}}^0=-|S|\left(\frac{\mu_1^2}{2}+\frac{\mu_2^2}{2}\right)
<\bar{\mathcal{E}}_{\mathrm{s}} \,.
\end{equation}
If $\alpha$ increases, the energy of the mixed configuration
grows, until it reaches a local maximum at $\alpha^{\mathrm{l}}$, which
marks the beginning of the nonphysical region. At this point the
density of the second species vanishes, and thus the energies of
the mixed and separated configurations are equal: they are in
fact both single-condensate configurations, whose energy is
\begin{equation}
\mathcal{E}_{\mathrm{s}}^{\mathrm{l}}=-|S|\frac{\mu_1^2}{2}=
\bar{\mathcal{E}}_{\mathrm{s}} \,.
\end{equation}
For $\alpha^{\mathrm{l}}<\alpha<\alpha^{\mathrm{u}}$, the
separated ones (not only the minimizing one) are the only
configurations that have a physical meaning. If
$\alpha=\alpha^{\mathrm{u}}$, the mixed configurations become
physical again, but their energy
$\mathcal{E}_{\mathrm{m}}^{\mathrm{u}}$ is higher than that of the
single-condensate configuration
\begin{equation}
\mathcal{E}_{\mathrm{m}}^{\mathrm{u}}=-|S|\frac{\mu_2^2}{2}>
\bar{\mathcal{E}}_{\mathrm{s}} \,.
\end{equation}
The single-condensate configuration remains energetically favored
for $\alpha \to \infty$, since its energy is constant, while
the energy of the mixed configuration (\ref{engcM}) vanishes as
$\alpha^{-1}$.

We now consider the case $\mu_2=\mu_1$.
It was observed that this is the only case in which
real separated states minimize the grand-canonical energy
$\mathcal{E}_{\mathrm{s}}$, and thus it is possible for such
configurations to be the ground state of the system. Moreover, we
stressed that if condition \eref{stat} holds, there is no
nonphysical region for the mixed configurations, as $\alpha$ varies
from zero to infinity. By plugging condition \eref{stat} in
\eref{engcM}, we find that the grand-canonical energy in this
case is never singular, and reads
\begin{equation}\label{engcL}
\mathcal{E}_{\mathrm{m}}=-|S|\frac{\mu_1^2 }{1+\alpha} .
\end{equation}
If \eref{engcL} is compared with (\ref{engcS}), that for $\mu_1=\mu_2$ reads
\begin{equation}
\mathcal{E}_{\mathrm{s}}=-|S|\frac{\mu_1^2}{2}
\end{equation}
and is independent of the  partition, mixed
configurations are found to be favorite if
$\alpha<1$, while separated configurations have
smaller energy if $\alpha>1$. Thus, even in the
case of fixed chemical potentials and infinite square well
external potential, the value \eref{u12lim} proves to be the
discriminant value between mixed and separated ground states.

\subsubsection{Generic potential.}

Consider TF density profiles $\rho_k(x)$. In $S_{11}=S_1-S_2$ we get (with the convention $N_1=N_2=0$)
\begin{equation}
\fl \qquad \mathcal{E}_{\mathrm{s}}^{(1)} = \frac{1}{2} \int_{S_{11}}  (\mu_1- V_1)^2 \, dx
- \int_{S_{11}}  (\mu_1- V_1)^2 \, dx = - \frac{1}{2} \int_{S_{11}}  (\mu_1- V_1)^2 \, dx,
\end{equation}
and in $S_{22}=S_2-S_1$
\begin{equation}
\mathcal{E}_{\mathrm{s}}^{(2)} = - \frac{1}{2} \int_{S_{22}}  (\mu_2- V_2)^2 \, dx .
\end{equation}
Therefore
\begin{equation}
\mathcal{E}_{\mathrm{s}}= - \frac{1}{2} \int_{S_{11}}  \tilde{V}_1(x)^2\, dx - \frac{1}{2} \int_{S_{22}}  \tilde{V}_2(x)^2\, dx ,
\end{equation}
with
\begin{equation}
\tilde{V}_k (x) = \mu_k - V_k(x), \qquad (k=1,2).
\end{equation}
In $S_{12}=S_1\cap S_2$ we get
\begin{equation}
\mathcal{E}_{\mathrm{m}}
= - \frac{1}{2(1-\alpha^2)} \int_{S_{12}}
\left( \tilde{V}_1(x)^2 + \tilde{V}_2(x)^2 -2 \alpha \tilde{V}_1(x)
\tilde{V}_2(x) \right)\, dx .
\end{equation}
Compare a mixed TF configuration in a set $S$ with a configuration with only one species, say
$\rho_1$,
\begin{eqnarray}
\delta \mathcal{E} &=& \mathcal{E}_{\mathrm{s}}^{(1)}-
\mathcal{E}_{\mathrm{m}} = \frac{1}{2}  \int_S \tilde{V}_1^2\, dx
+ \frac{1}{2(1-\alpha^2)} \int_S \left(\tilde{V}_1^2
+\tilde{V}_2^2 -2 \alpha \tilde{V}_1 \tilde{V}_2 \right)\, dx +
\nonumber\\
& = & - \frac{1}{2(1-\alpha^2)} \int_S \left( \tilde{V}_2 - \alpha
\tilde{V}_1 \right)^2 \, dx.
\end{eqnarray}
Thus, when $\alpha>1$ one gets $\delta \mathcal{E} \leq 0$ and separated configurations are energetically favorite.
 Therefore, in this condition, we have to search for the ground state among the separated
configurations, which is the aim of the next
section.

\subsection{Remarks}
From the results obtained in this Section, it clearly emerges
that, if $\alpha \geq 1$, that is $U_{12}\geq U_{12}^{\mathrm{th}}$, the ground state is in a
separated configuration.
Moreover, according to the selection principle introduced in Sec.~\ref{sec:selprinc},
the following analysis will be  restricted to piecewise continuously differentiable
solutions  of the TF equations, $\rho_k\in\tilde{C}^1$,  whose discontinuities can be due
to the presence of a finite number of interfaces.

\section{Minimizing separated configurations}\label{sep-opt}

In \Sref{mix-sep} it was shown that separated configurations are
energetically favored if
\begin{equation}
\label{u12high} \alpha >1 \,,
\end{equation}
for generic continuously differentiable confining  potentials.
We will now find a way to determine which of
these configurations is the ground state of the system, and which
can be regarded to be locally stable or unstable. We shall again discuss separately the cases
of fixed numbers of particles and fixed chemical potentials,
underlining analogies and differences between them.

When only piecewise continuously differentiable density profiles
are considered, a separated configuration can be characterized by
the property that the supports $S_1$ and $S_2$ do not intersect,
except at a finite set of points. The intersection
points correspond to a set of domain walls separating the
first and the second species. $S_1$ and $S_2$, being compact, are thus unions of
intervals, which can be bounded by i) two domain walls, ii) a
domain wall and a zero of the TF density profile, or iii)  two
zeros. From \eref{TFA1}-\eref{TFA2}, the possible zeros
$\zeta_j^{(k)}$, with $j=1,\dots,m$  and $k=1,2$, of the densities $\rho_k$ are subject to the condition
\begin{equation}\label{zeros}
\mu_{k}=V_{k}(\zeta_j^{(k)}) \,.
\end{equation}
In general, if we search the ground state among the separated
configurations, we have to deal with the minimization of the sum of two decoupled
functionals of the kind
\begin{equation}
\mathcal{U}\left(\rho_1,\rho_2\right)=\mathcal{V}\left(\rho_1\right) + \mathcal{V}\left(\rho_2\right).
\end{equation}
We remark that, except for condition \eref{u12high}, that enables one
to establish that the ground state is a separated configuration,
the parameter $\alpha$ plays no role in the search for minimal
separated configurations.

We will proceed by fixing the number $n$ of domain walls and
determine the set of positions $\vec{R}=(R_1, \dots, R_n)$ for which the
considered functional, internal or grand-canonical energy, is, at
least locally, minimized. We will then compare the minima
corresponding to different numbers of walls.

\subsection{Fixed numbers of particles}

In this case our aim is to find the stationary configurations of the
internal energy
\begin{eqnarray}\label{enintW}
\fl \qquad \mathcal{U}\left(\rho_1,\rho_2\right)= \frac{1}{2} \int_{S_1}  \left( \rho_1(x)^2 + 2
V_1(x)\rho_1(x)\right)  dx +\frac{1}{2} \int_{S_2}  \left( \rho_2(x)^2 + 2
V_2(x)\rho_2(x)\right)  dx
\end{eqnarray}
with respect to small variations of $\vec{R}$, under
the condition that the numbers of particles $N_k$
remain fixed, $\int_{S_k}\rho_k \, dx=N_k$.
Then, conditions will be set for these stationary
configurations to be local minima. It can be easily inferred that
the chemical potentials in \eref{TFA1}-\eref{TFA2}, which are used
as Lagrange multipliers to normalize the density profiles, and
thus depend on the supports $S_1$ and $S_2$, are functions of the
domain wall positions $\vec{R}$. We re-express the separated TF
density profiles, explicitly showing this additional dependence:
\begin{eqnarray}\label{TFA1W}
\rho_k(x;\vec{R})=\mu_k(\vec{R})-V_k(x) \,, \qquad (k=1,2).
\end{eqnarray}
The zeros $\zeta_j^{(k)}$, subject to condition \eref{zeros},
are also functions of $\vec{R}$. Since the choice of the position
of the domain walls completely defines the density profiles and
their supports, the internal energy \eref{enintW}, evaluated at stationary TF densities, can be
viewed as a function of $\vec{R}$
\begin{equation}\label{enintWf}
\mathcal{U}\left(\rho_1(\cdot; \vec{R}),\rho_2(\cdot;\vec{R})\right):=U(\vec{R}) \,.
\end{equation}
We now assign to each domain wall a dichotomic variable $s_j$:
$s_j=+1$ if it is the upper border of a interval containing the
first species (and thus the lower border of an interval containing
the second one), and $-1$ in the complementary case.

By taking the first derivative of \eref{enintWf} with respect to a
generic $R_j$ and using normalization conditions, one gets
\begin{equation}\label{enintder}
\frac{\partial U (\vec{R})}{\partial R_j} =\frac{s_j}{2}\left(
\rho_2(R_j;\vec{R})^2-\rho_1(R_j;\vec{R})^2\right)
\,.
\end{equation}
The stationarity condition is obtained by setting to zero the
derivatives \eref{enintder} with respect to all the positions of
the $n$ domain walls, yielding
\begin{equation}\label{densratioW}
\rho_1(R_j;\vec{R})=\rho_2(R_j;\vec{R})
\quad\forall j=1,\dots\,n \,.
\end{equation}
These are clearly analogous to the minimum conditions
\eref{densratio} in the case of an infinite potential well. If the
densities in \eref{densratioW} are expressed as functions of the
external potentials and the chemical potentials, it becomes
clear that the position of the domain walls in a stationary
configuration are characterized by the fact that the potential
\begin{equation}
\varphi(x):=V_1(x)-V_2(x)
\end{equation}
is equal for all $R_j$:
\begin{equation}
\varphi(R_j)
=\mu_1(\vec{R})-\mu_2(\vec{R}).
\end{equation}
Let us consider the equation
\begin{equation}
\varphi(x)=f,
\end{equation}
with $f$ a constant. The number of its solutions fixes the maximal
number of walls in a stationary configuration.

It is particularly interesting the case in which the external
potentials for the two species are proportional,
\begin{equation}
\label{propV}
V_2(x)=\beta V_1(x) =: \beta V(x)
\label{eq:propV}
\end{equation}
with $\beta>0$. This happens, e.g.\ when the two condensates feel the same potential before the scaling~(\ref{eq:scaling}), and in such a situation
\begin{equation}
\label{betadef}
\beta=\sqrt{\frac{U_{11}}{U_{22}}}.
\end{equation}
In the case of proportional potentials (\ref{propV}), if the
equation $V(x)=v$, with $v$ a constant, has $n$ solutions, there
cannot exist stationary configurations with more than $n$ domain
walls. In this case, the  domain walls are placed at positions characterized by the same potential,
which must be equal to
\begin{equation}\label{eqV}
V(R_j)=\frac{\mu_1(\vec{R})-\mu_2(\vec{R})}
{1-\beta}, \qquad \forall j=1,\dots,n \,.
\end{equation}
As a consequence, the densities of the same species must be equal
at the edge of each domain wall. The values of the densities at
the edge of all domain walls in the case of proportional
potentials will be indicated as
$\tilde{\rho}=\rho_1(R_j)=\rho_2(R_j)$.

\begin{figure}
\includegraphics[width=\textwidth]{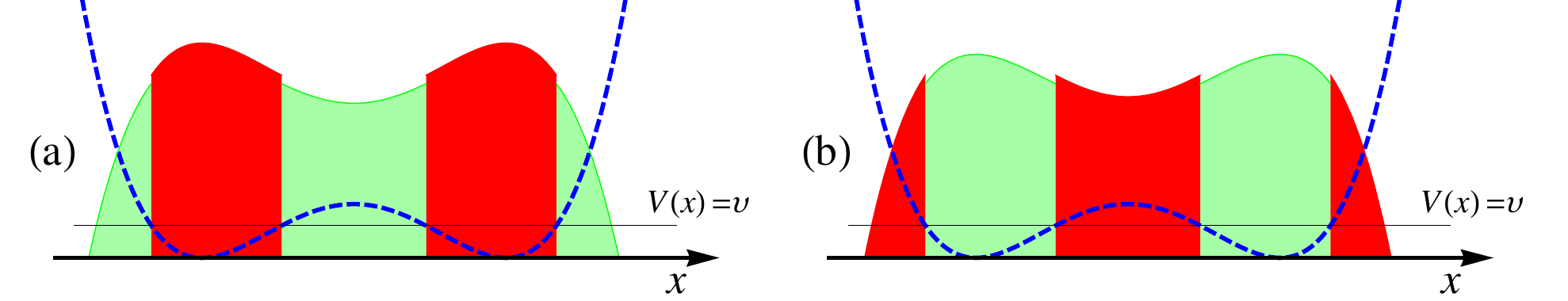}
\caption{(Color online). Examples of maximal stationary
configurations in a symmetric double-well potential. The blue
dotted lines represent the external potential, in arbitrary units.
The density profile of the first species is in dark grey (red in
the online version), while the density profile of the second one
is in light grey (green in the online version). In (a) the minima
of the potential are occupied by the first species, which is the
less self-interacting one. In (b) the minima are occupied by the
second species. Computation of the second derivatives shows that
only configuration (a) is (at least locally)
stable.}\label{maximal}
\end{figure}

Henceforth, we shall call \textit{maximal stationary
configurations} those ones in which a domain wall is placed in each of
the real solutions of \eref{eqV}, except for the case in which one
of the solutions is a stationary point for $V(x)$. Two different
examples of such configurations are shown in \Fref{maximal}.
In the following, it will be shown that for $\beta\approx1$ the ground state of the
system is  usually in a maximal configuration.

In order to determine if the stationary configurations are in fact
minima of \eref{enintWf}, the Hessian matrix $H$ at the stationary
solution has to be computed. By deriving \eref{enintder} once more
with respect to $R_j$, we find that $\partial^2 U/\partial R_j^2$
has two contributions: the first one is related to the
dependence of the external potentials on the point $R_j$, the
second one to the dependence of the chemical potentials on the
position of the domain walls $\vec{R}$. On the other hand, if
\eref{enintder} is derived with respect to $R_k$ with $j\neq k$,
only the second one of the above mentioned contributions survives.
By taking into account conditions \eref{densratioW}, the diagonal
elements of the Hessian matrix in the stationary configurations
reads
\begin{eqnarray}\label{hessdiag}
H_{jj}= s_j\rho_1(R_j;\vec{R})
\varphi'(R_j)
+\left( \frac{1}{|S_1|}
+\frac{1}{|S_2|} \right)\rho_1(R_j;\vec{R})^2 \,,
\end{eqnarray}
while the non diagonal elements are
\begin{equation}
H_{jk}=s_js_k\left( \frac{1}{|S_1|} +\frac{1}{|S_2|}
\right)\rho_1(R_j;\vec{R}) \rho_1(R_k;\vec{R}) \,.
\end{equation}
For large numbers of particles, so that $|S_k|$ are sufficiently large, all terms depending on
the inverse length can be neglected, and conditions for the Hessian
matrix to be positive definite, and hence for the corresponding
stationary configuration to be a local minimum, are easy to find:
\begin{eqnarray}
\varphi'(R_j)>0 \quad \Leftrightarrow \quad
 V'_1(R_j) > V'_2(R_j) \qquad \mathrm{if }\quad s_j=+1 \,,
\nonumber \\
\varphi'(R_j)<0 \quad \Leftrightarrow \quad
 V'_1(R_j) < V'_2(R_j) \qquad \mathrm{if }\quad s_j=-1 \,.
\label{hesspos2}
\end{eqnarray}
For smaller numbers of particles the complete Hessian matrix has to be
diagonalized (e.g.\ in a numerical way). A simpler and relevant
case is that of equal external potentials, in which we have
already remarked that in a stationary configuration the values of the densities of each
species must be equal at all the domain walls. We obtain
\begin{equation}\label{hessianep}
H_{jk}^{\mathrm{pp}}= \delta_{jk}a_j  +(-1)^{j+k}C \,,
\end{equation}
where we have taken into account that $s_js_k=(-1)^{j+k}$ and
defined
\begin{equation}\label{aj}
a_j=s_j\tilde{\rho}
\left(1-\beta\right) V'(R_j)
\end{equation}
as the intensive and purely diagonal part, and
\begin{equation}
C=\left( \frac{1}{|S_1|}+\frac{1}{|S_2|}
\right)\tilde{\rho}^2
\end{equation}
as the lenght-dependent term, which vanishes for large numbers of particles. In this limit, when $\beta<1$ the
condition for a stationary configuration to be locally stable is
that the potential $V(x)$ be increasing at \textit{all} the
$R_j$'s which are the upper (right) border of an interval
containing particles of the first species ($s_j=+1$), and
decreasing at \textit{all} the $R_j$'s which are the lower (left)
border of an interval of the same kind ($s_j=-1$).
Intuitively, the less self-interacting condensate [since we
 supposed that $\beta<1$ in Eq.\ (\ref{betadef})] tends to occupy regions of the
real axis in which the potential is lower, while the most
self-interacting one is pushed into regions where the external
potential is higher. For small numbers of particles, such
configurations continue to be minima for \eref{enintWf}, since if
the potential is increasing when $s_j=+1$ and decreasing when
$s_j=-1$, $H_{jk}^{\mathrm{pp}}$ in Eq.\
\eref{hessianep} is the sum of two positive definite
matrices, and hence it is positive definite. Moreover, it is
possible that even a configuration in which
\begin{equation}
a_j\leq 0 \qquad \mathrm{for}\quad\mathrm{some}\quad
j\in\{1,\dots,n\}
\end{equation}
becomes stable, which is impossible in the thermodynamic limit.
However, bounds on the stability of such configurations can be
found if some necessary conditions for \eref{hessianep} to be
positive definite are tested. First, since
$|H_{ij}^{\mathrm{pp}}|\leq(H_{ii}^{\mathrm{pp}}+H_{jj}^{\mathrm{pp}})/2$
for all pairs of indices, then
\begin{equation}
a_i+a_j\geq 0 , \qquad\forall i,j=1,\dots,n ,
\end{equation}
which implies that only one of the $\{a_j\}$, say $a_{\bar{j}}$,
can be nonpositive for a stable configuration, and moreover
\begin{equation}
|a_{\bar{j}}|\leq\min_{j\neq \bar{j}}a_j .
\end{equation}
On the other hand, by applying on the other hand the necessary condition $\det H>0$,
another constraint can be established,
\begin{equation}
|a_{\bar{j}}| <\frac{C}{1+C\sum_{j\neq \bar{j}}a_j^{-1}}  ,
\end{equation}
with the upper bound vanishing in the thermodynamic limit.

\begin{figure}
\includegraphics[width=\textwidth]{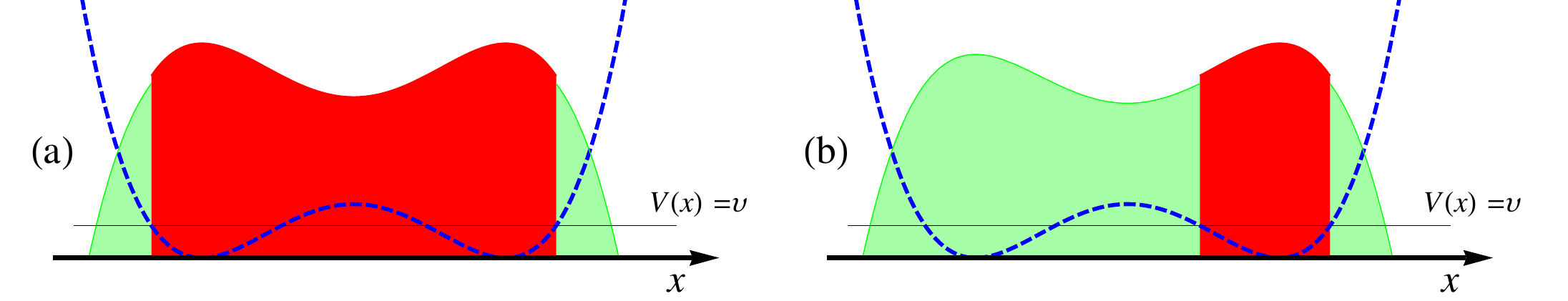}
\caption{(Color online). Non-maximal configurations in a
double-well potential. Lines and shades have the same meaning as
in \Fref{maximal}. In (a) the first species occupies a region
around the potential barrier, where the potential is higher than
the value $v$ in the domain walls. In (b) the second species
occupies one of the minima of the external potential, where it is
lower than the value $v$ in the domain walls.}\label{nonmax}
\end{figure}

There are two kinds of maximal stationary configurations: the first one is characterized by the fact
that the external potential at each point of $S_1$ is smaller than
the potential at each point of $S_2$, while the second one is
characterized by the opposite situation. The latter, however, is
not stable, since the diagonal part of its Hessian matrix contains
$a_i<0$ for all $i=1,\dots,n$. We can thus limit
our attention to the first kind of profiles, which we call the
\textit{maximal stable configurations} (see \Fref{maximal}). We
will now prove that if a locally stable configuration is not
maximal, there are conditions ensuring that it cannot be the
ground state of the system. The proof is based on the fact that in
a non-maximal and locally stable configuration one of the following
situations emerges: either the potential at a point of $S_1$ is
greater than the potential at the domain walls, or the potential
at a point of $S_2$ is smaller than its value at the domain walls.
Examples of non-maximal configurations are represented in
\Fref{nonmax}.
We remark that both these situations can be present in the same
configuration. We start by considering the first one. Let
\begin{equation}
v= V(R_j)
\end{equation}
be the potential at the domain walls. Assume that the
potential at a point $x_0\in S_1$ be such that
\begin{equation}
V(x_0)=: \bar{V}_1>v \,.
\end{equation}
We exclude the case $\bar{V}_1=\mu_1$, implying a vanishing density at
$x_0$. (This case will be eventually considered as a limit.) We
now consider a subinterval $\omega_{\epsilon}$  of $S_2$ of length
$\epsilon>0$, with a domain wall as one of its edges, and
an interval $\omega_{\eta}$ of length $\eta>0$, which is a
neighborhood of $x_0$, and impose that the number of first-species particles in
$\omega_{\eta}$ be equal to $n_0$ and that of second-species
particles in $\omega_{\epsilon}$ be $n_0/\beta$. Since the potential is regular
and the densities are supposed to be regular between each pair
of domain walls, one gets
\begin{equation}
n_0=\eta\rho_1(x_0)+\Or(\eta^2)=\epsilon \beta \rho_2(R_j)+\Or(\epsilon^2)
\,.
\end{equation}
This equality implies a relation between the length of the
considered intervals, depending on the ratio of the densities,
which can be expressed in terms of the potentials by using
condition \eref{densratioW}:
\begin{equation}\label{epseta}
\eta(\epsilon)=\epsilon \beta \frac{\rho_2(R_j)}{\rho_1(x_0)}+\Or(\epsilon^2)
=\epsilon
\beta \frac{\mu_1-v}{\mu_1-\bar{V}_1}+\Or(\epsilon^2)
\,.
\end{equation}
The potential energy of the two selected intervals is given by the
sum of the contributions
\begin{eqnarray}
u_{\epsilon}^{(2)}=
v n_0+\frac{1}{2}\frac{n_0^2}{\epsilon\beta^2}+\Or(\epsilon^2) \,,
\\
u_{\eta}^{(1)}=
\bar{V}_1n_0+\frac{1}{2}\frac{n_0^2}{\eta(\epsilon)}+\Or(\epsilon^2)
\,.
\end{eqnarray}
We now replace the original density profiles with flat density
profiles which preserve the numbers of particles. In particular,
we fill $\omega_{\epsilon}$, which initially belonged to $S_2$, with the
first-species condensate with a density
\begin{equation}
\bar{\rho}_1=\frac{n_0}{\epsilon} \,,
\end{equation}
and $\omega_{\eta}$ with the second-species particles with a density
\begin{equation}
\bar{\rho}_2=\frac{n_0}{\eta(\epsilon) \beta} \,.
\end{equation}
With these new density profiles, the internal energy of the
intervals becomes the sum of the terms
\begin{eqnarray}
u_{\epsilon}^{(1)}=
vn_0+\frac{1}{2}\frac{n_0^2}{\epsilon}+\Or(\epsilon^2) \,, \\
u_{\eta}^{(2)}=
 \bar{V}_1n_0+\frac{1}{2}\frac{n_0^2}{\eta(\epsilon) \beta^2}+\Or(\epsilon^2)
\,.
\end{eqnarray}
The total variation of the internal energy induced by this change
is
\begin{equation}\label{deltaU}
\delta\mathcal{U}= u_{\epsilon}^{(1)}+u_{\eta}^{(2)}
-u_{\epsilon}^{(2)}-u_{\eta}^{(1)}=
\frac{n_0^2}{2}\left(\frac{1}{\beta^2}-1\right)
\left(\frac{1}{\eta(\epsilon)}-\frac{1}{\epsilon}
\right)
+\Or(\epsilon^2) .
\end{equation}
Since $\beta<1$, we find that if $\eta(\epsilon)>\epsilon$,
i.e., up to first order in $\epsilon$,
\begin{equation}\label{nonmax1}
 \frac{\mu_1-v}{\mu_1-\bar{V}_1}>\frac{1}{\beta} \,,
\end{equation}
and it is  always possible, for sufficiently small $\epsilon$,
to find a density profile that preserves the numbers of particles,
whose energy is smaller than the energy of a non-maximal stable
configuration. It can be observed that if the limit
$\bar{V}_1\rightarrow\mu_1$ is taken, condition \eref{nonmax1} is certainly
satisfied. This means that in the ground-state configuration the
intervals of the support $S_1$ of the less self-interacting species
cannot be bordered by a zero.

If the case in which there exists a point in $x_0\in S_2$ where the
external potential is lower than its value in the domain walls,
\begin{equation}
V(x_0)=:\bar{V}_2<v \,,
\end{equation}
we find, with the same procedure as in the previous case, that if the following inequality is satisfied
\begin{equation}\label{nonmax2}
\frac{\mu_2-v}{\mu_2-\bar{V}_2}<\beta \,,
\end{equation}
the considered non-maximal configuration can never be the ground
state of the system. Since usually $\beta^2= U_{11}/U_{22}\simeq
1$ (see, e.g., the hyperfine states of $^{87}\mathrm{Rb}$,
\cite{kasamatsu}), conditions \eref{nonmax1}-\eref{nonmax2} set a
very stringent limitation on the possibility that a non-maximal stable
configuration be the ground state of the binary mixture.

\subsection{Fixed chemical potentials}

The results in the case of fixed chemical potentials is very
similar to that of fixed numbers of particles, in the
thermodynamical limit. The functional to be minimized by separated
configuration is (we set $N_1=N_2=0$)
\begin{equation}\label{engcW}
\fl \qquad \mathcal{E}\left(\rho_1,\rho_2\right)=  \int_{S_1}  \left(\frac{1}{2}\rho_1^2+
V_1 \rho_1-\mu_1\rho_1\right)dx
+\int_{S_2} \left(\frac{1}{2}\rho_2^2+
V_2 \rho_2 -\mu_2\rho_2\right)  dx
\,.
\end{equation}
An important difference with respect to the previous case is that,
since the chemical potentials are fixed and not subject to
normalization conditions, the TF density functions are completely
independent of the positions of the domain walls. Thus, the
functional \eref{engcW} depends on $\vec{R}$ only through the
domains of integrations, which are determined by the supports of
the density profiles, and it can be seen again as a function of
the domain wall positions:
\begin{equation}\label{engcWf}
\mathcal{E}\left(\rho_1(\cdot;\vec{R}),\rho_2(\cdot;\vec{R})\right):=E(\vec{R}) \,.
\end{equation}
The stationarity conditions are exactly the same as in the case of
fixed numbers of particles, since the first derivative with
respect to a generic $R_j$ reads
\begin{equation}\label{engcder}
\frac{\partial E(\vec{R})}{\partial R_j}=\frac{s_j}{2}\left(
\rho_2(R_j;\vec{R})^2-\rho_1(R_j;\vec{R})^2\right)
\,.
\end{equation}
However, the Hessian matrix in the stationary configurations
is diagonal, as in \eref{hessdiag}. The
absence of the non-diagonal part lies in the fact that the first
derivative \eref{engcder} depends on the position of the domain
walls only through the external potentials. The
stability criterions for a stationary density profile are the same
as in the case of fixed numbers of particle, if the
thermodynamical limit is considered: a configuration is stable if
and only if conditions \eref{hesspos2} are
satisfied for all $j$. If the two species lie in the same external
potential, these conditions reduce to $s_j V'(R_j)>0$.

Even in the case of fixed chemical potentials, it is possible to
show that if the potentials are proportional (\ref{eq:propV}),
there are limitations on the possibility that a non-maximal stable
configuration be the ground state. Indeed, if at the domain walls
we have $V(R_j)=v$, it can be shown that if there exist a point in
$S_1$, where the potential is $\bar{V}_1>v$ and satisfies
\begin{equation}\label{nonmax1CP}
\frac{\mu_2-\bar{V}_1}{\mu_1-\bar{V}_1}>\frac{1}{\beta}
\,,
\end{equation}
then the grand-canonical energy of the configuration is higher
than the energy of another configuration corresponding to the same
chemical potentials. We observe that the stationarity condition
\eref{densratioW} (with the chemical potentials
\textit{independent} of $\vec{R}$) implies that the chemical
potential $\mu_2$ be greater than $\mu_1$. Thus, the left hand
side of \eref{nonmax1CP} is an increasing function of $\bar{V}_1$.
Moreover, the equality is saturated for $\bar{V}_1=v$. We remark
that if the limit $\bar{V}_1\rightarrow \mu_1$ is considered, that
is if the density profile of the first species condensate has a
zero, condition \eref{nonmax1CP} is certainly verified, and then
such a configuration cannot be the ground state of the system. On
the other hand, if there exist a point in $S_2$ where the
potential is $\bar{V}_2<v$, satisfying
\begin{equation}\label{nonmax2CP}
\frac{\mu_1-\bar{V}_2}{\mu_2-\bar{V}_2}>\beta
\,,
\end{equation}
then there exist another configuration corresponding to the same
chemical potentials, which has a lower grand-canonical energy. In
this second case, the left hand side of \eref{nonmax2CP} is a
decreasing function of $\bar{V}_2$, and the equality is again
satisfied by $\bar{V}_2=v$.

Here, we sketch the proof in the first case, the second case being
analogous. Let us suppose that at a given point $x_0\in S_1$ the
potential satisfies $V(x_0)=\bar{V}_1>v$, and consider two
intervals $\omega$ and $\chi$ of the same length $\epsilon$, $\omega$ lying in $S_2$
and bordered by a domain wall, and $\chi$ lying in $S_1$ and
containing $x_0$. The chemical potentials are fixed, thus the
functional form of the density functions does not depend on the
positions of the domain walls. We now replace in $\omega$ the
first species with the second one, and in $\chi$ the second
species with the first one, using again TF density functions.
Taking into account condition $\eref{densratioW}$, we find that
the difference between the energy of the final and the initial
configurations is
\begin{equation}
\delta\mathcal{E}=\epsilon\frac{(\mu_1-\bar{V}_1)^2}{2}
\left[1-\left(\beta\frac{\mu_2-\bar{V}_1}
{\mu_1-\bar{V}_1}\right)^2\right]+\Or(\epsilon^2) \,.
\end{equation}
For sufficiently small $\epsilon$, the final configuration is
energetically favored with respect to the initial one, if
condition \eref{nonmax1CP} applies. We finally note that, unlike in
\eref{nonmax1}-\eref{nonmax2}, the conditions
\eref{nonmax1CP}-\eref{nonmax2CP} are independent of the value of
$v$. Even in this case, since usually $\beta^2=U_{11}/U_{22}\simeq 1$,
conditions \eref{nonmax1CP}-\eref{nonmax2CP} preclude any
non-maximal stable configuration from being the ground state.

\section{Conclusion}\label{concl}

We have studied the
Thomas-Fermi equations, for a system of two Bose-Einstein
condensates confined in generic potentials. We have emphasized the
role of the limiting value $U_{12}=\sqrt{U_{11}U_{22}}$ in
determining if the ground state of the system is a mixed
configuration or a stationary one, by assuming that the external
potentials be  regular. We then determined a set of conditions to be
satisfied by locally stable separated configurations. Then we
looked for the ground state among the possible locally stable
configurations, and found that those with a maximal numbers of
domain walls are usually energetically favorite. The results
presented in this article enable us to find the ground state of
binary mixtures in multi-well potentials, given either the numbers
of particles or the chemical potentials.

It would be interesting to analyze the changes that a correction
to the TF approximation, including the kinetic energies, would introduce in such a
picture. If the numbers of particles are sufficiently high, the
TF approximation is very accurate. Nonetheless, TF density profiles correspond
to diverging kinetic energy, due in particular to the
discontinuities at the domain walls. The kinetic parts intervenes
by regularizing the TF solutions, at the expense of an increase
in the potential energy, especially in a neighborhood of a domain
wall. This could lead to an inversion in the energetic diagram, in
which configurations with few domain walls could become
energetically favorite with respect to maximal stable
configurations. This inversion has already been numerically
studied in the simple case of a harmonic potential
\cite{kasamatsu}, but the tools introduced in this article
uncover the possibility of extending this kind of analysis to
generic multi-well potentials, such as arrays of optical traps,
which are now within experimental reach.

\end{document}